\documentclass[11pt,twoside]{article}
\usepackage{./asp2014}

\aspSuppressVolSlug
\resetcounters

\begin{document}

\title{Galaxy And Mass Assembly (GAMA): A study of energy, mass, and structure (1kpc-1Mpc) at $z<0.3$}
\author{Simon P. Driver,$^{1,2}$ (and the GAMA team)}
\affil{$^1$International Centre for Radio Astronomy Research (ICRAR), School of Physics, University of Western Australia, Crawley, Perth, WA 6009, Australia; \email{Simon.Driver@uwa.edu.au}}
\affil{$^2$SUPA, School of Physics and Astronomy, University of St Andrews, North Haugh, St Andrews, Fife, KY16 9SS, UK, \email{spd3@st-and.ac.uk}}

\paperauthor{Simon P.Driver}{Simon.Driver@uwa.edu.au}{}{International Centre for Radio Astronomy Research}{University of Western Australia}{Crawley}{Perth}{WA 6009}{Australia}

\begin{abstract}
The GAMA survey has now completed its spectroscopic campaign of over
250,000 galaxies ($r<19.8$mag), and will shortly complete the
assimilation of the complementary panchromatic imaging data from
GALEX, VST, VISTA, WISE, and Herschel. In the coming years the GAMA
fields will be observed by the Australian Square Kilometer Array
Pathfinder allowing a complete study of the stellar, dust, and gas
mass constituents of galaxies within the low-z Universe ($z<0.3$).
The science directive is to study the distribution of mass, energy,
and structure on kpc-Mpc scales over a 3billion year timeline. This is
being pursued both as an empirical study in its own right, as well as
providing a benchmark resource against which the outputs from
numerical simulations can be compared. GAMA has three particularly
compelling aspects which set it apart: completeness, selection, and
panchromatic coverage. The very high redshift completeness ($\sim
98$\%) allows for extremely complete and robust pair and group
catalogues; the simple selection ($r<19.8$mag) minimises the selection
bias and simplifies its management; and the panchromatic coverage,
0.2$\mu$m - 1m, enables studies of the complete energy distributions
for individual galaxies, well defined sub-samples, and population
assembles (either directly or via stacking techniques). For further
details and data releases see: http://www.gama-survey.org
\end{abstract}

\section{Introduction}

\vspace{-0.5cm}

\noindent
Extra-galactic studies, in and around the 21st century, can arguably
be broken down into four distinct categories: Focused experiments
(e.g., WiggleZ, BOSS, DES, Euclid); high-fidelity studies of well
selected sub-samples (e.g., S$^4$G, ATLAS3D, MANGA etc); frontier
studies (e.g., HDF, UDF, Frontier's fields); and open-ended legacy
studies (e.g., 2MASS, SDSS, COSMOS, GEMS, CANDLES). These distinct
approaches are all important and highly complementary. The Galaxy And
Mass Assembly survey (GAMA; Driver et al.~2011), very much fits into
the latter category, by providing a broad legacy resource to the
community, with a key focus on being comprehensive and complete. GAMA,
like its predecessors the SDSS and 2MASS, is now forming the basis for
high-fidelity follow-on studies (e.g., SAMI/IFU, ASKAP/DINGO, Euclid
Legacy Science), and even frontier studies (e.g., HST lens sample,
JWST usage of GAMA groups as probes to $z>10$). Internally the GAMA
team now consists of over 100 scientists studying the distribution and
evolution of mass, energy, and structure with data also flowing through to
external teams fueling collaborative projects.

\articlefigure{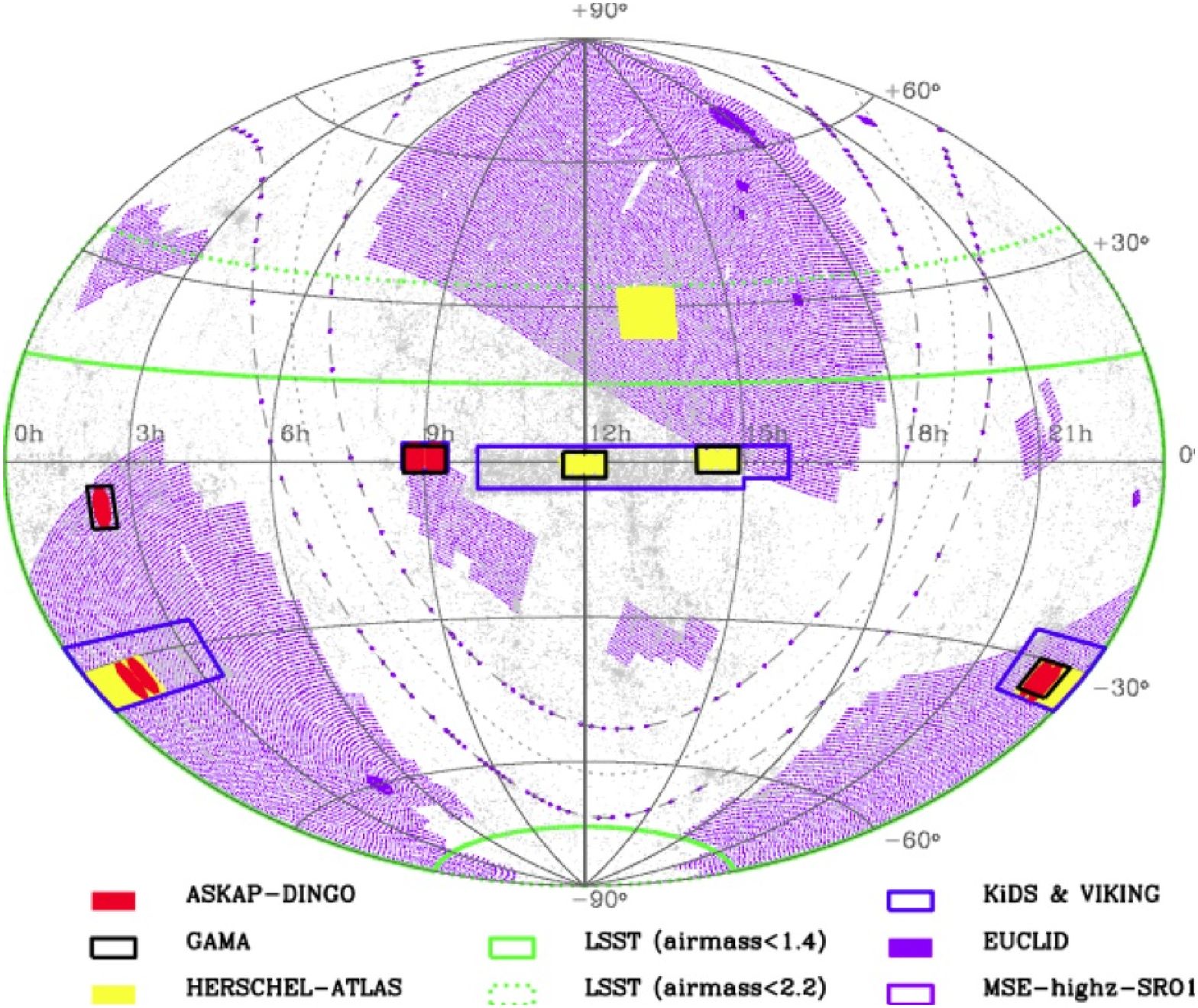}{SDriver-fig1x.eps}{An Aitoff projection showing the
  location of the GAMA fields with respect to other notable, ongoing,
  or planned surveys, in particular Herschel-Atlas, VST KiDS, ASKAP
  DINGO, Euclid and LSST (as indicated).}

\section{Galaxy And Mass Assembly}

\vspace{-0.5cm}

\noindent
The GAMA survey covers four 60 sq deg regions (3 equatorial, 1
southern), plus an additional 20 sq deg region (i.e., 260sq deg in
total), see Fig.~1 \& 2. The three primary equatorial blocks: G09, G12
and G15 lie within the SDSS Main survey providing the key input
catalogue to the spectroscopic campaign. The input cat for the G02
region is CFHTLS-W1 and the input cat for the southern block (G23)
is provided by VST KiDS. In due course all regions will be covered by
VST KiDS to comparable depth. Details on the equatorial sample
selection are provided in Baldry et al.~(2010) and essentially rely on
size and flux cuts augmented by near-IR colours used to recoup compact
systems.

The spectroscopic observations were conducted over a period of 210
nights spanning 5years using the AAOmega spectrographs coupled to the
2dF fibre-optic positioning system ($R \sim 300$, 400 fibres,
3700-9000\AA) mounted on the Anglo-Australia Telescope.  Details of
the observing strategy, spectroscopic reduction pipeline, and data
releases are described in Robotham et al.~(2010), Hopkins et
al.~(2013), Driver et al.~(2011) and Liske et al.~(2015). The current
catalogue comprises a sample of over 250,000 galaxies with $\sim 98$\%
spectroscopic completeness (r.f., SDSS, 85\% and zCOSMOS,
60\%). Fig.~2 highlights the portion of the Universe surveyed by GAMA,
extending 5Gyrs outwards in lookback time in the five distinct
regions. Also shown on Fig.~2 are the other surveys one can consider
as complete (i.e., unbiased selections).

\articlefigure{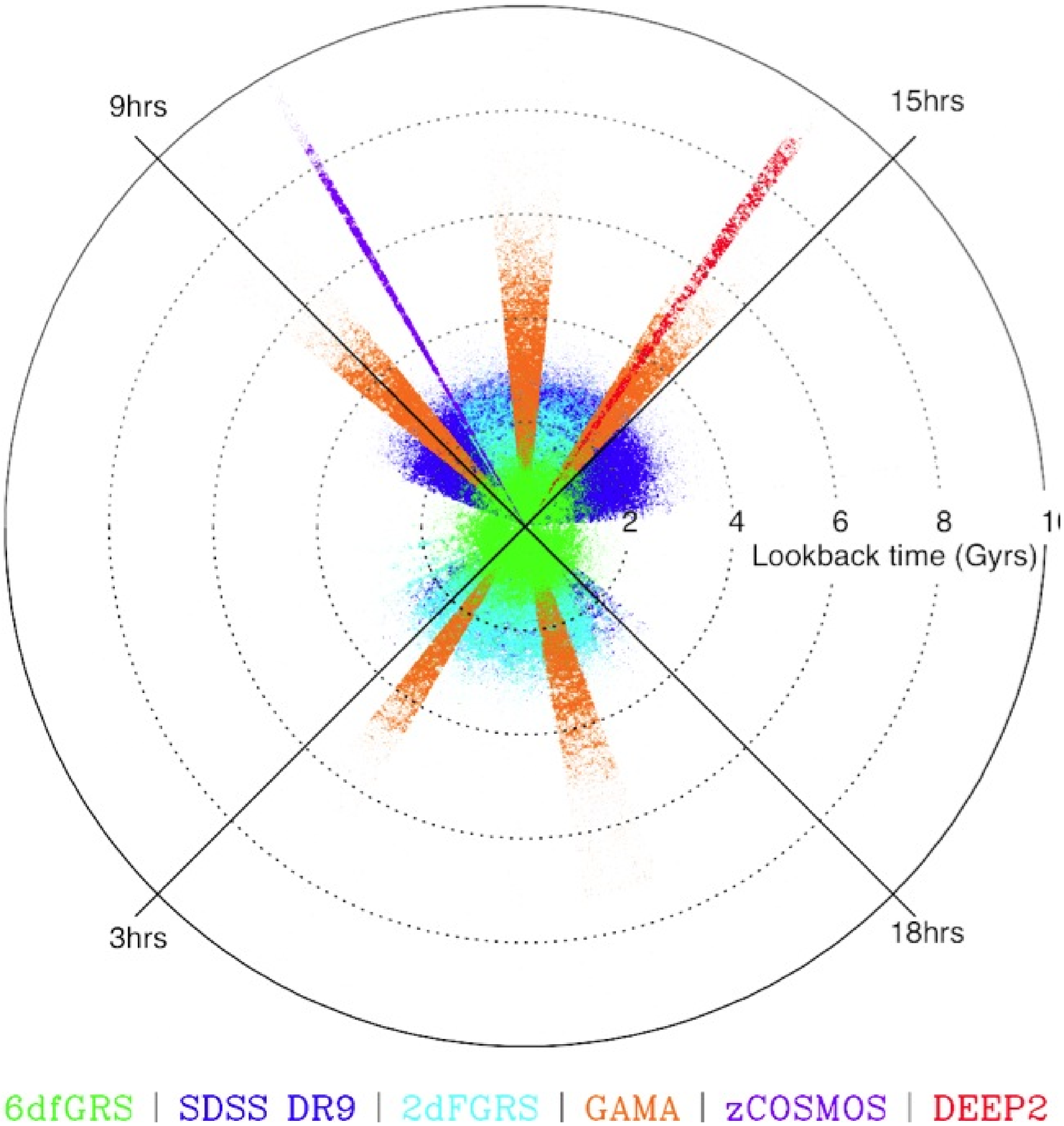}{SDriver-fig2x.eps}{Cylindrical plots of various
  complete galaxy surveys as indicated.}

\articlefigure{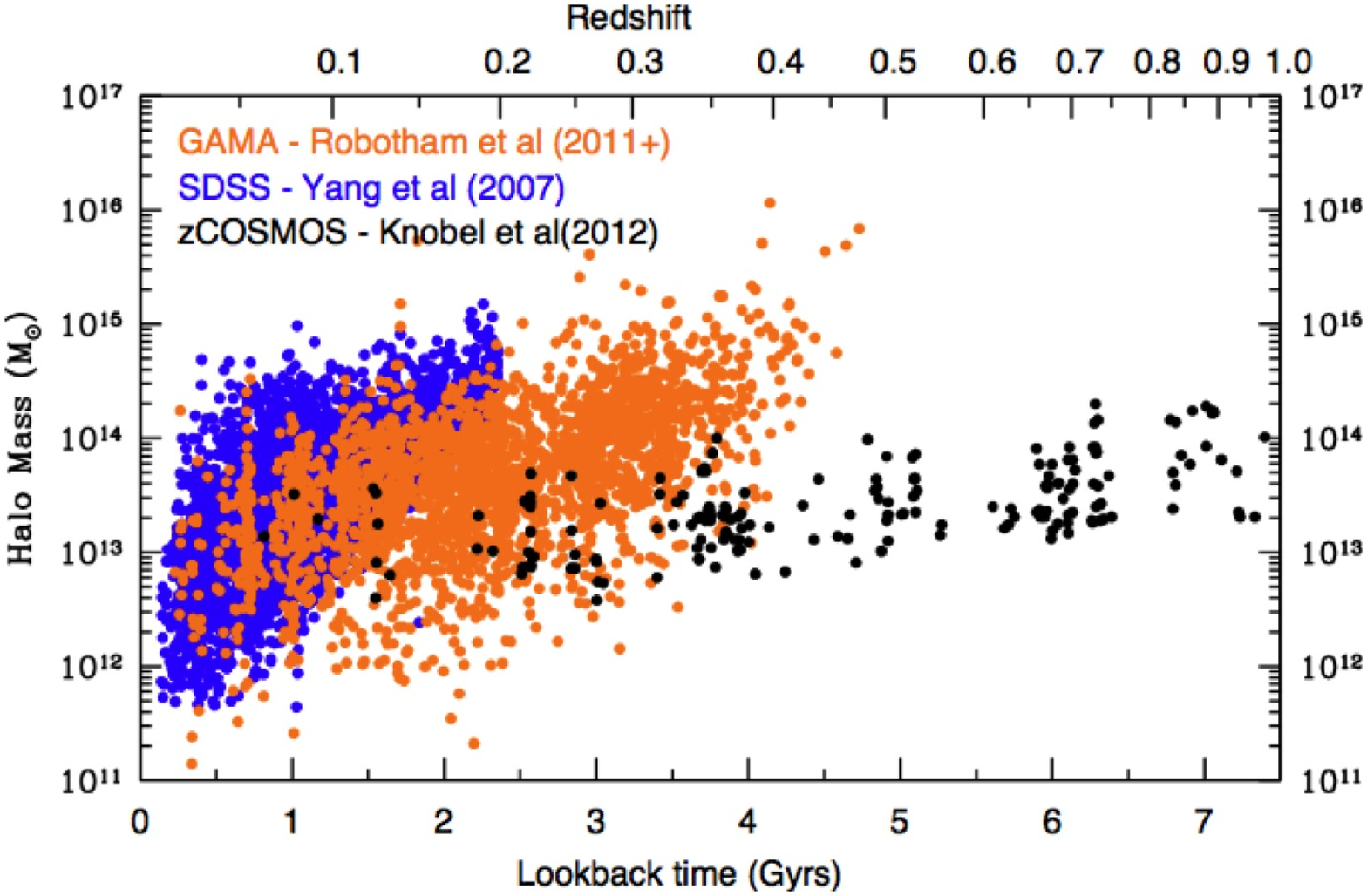}{SDriver-fig3x.eps}{A
  comparison of the halo-mass lookback space explored by the three
  notable group catalogues derived from SDSS, GAMA and zCOSMOS (as
  indicated).}

\articlefigure{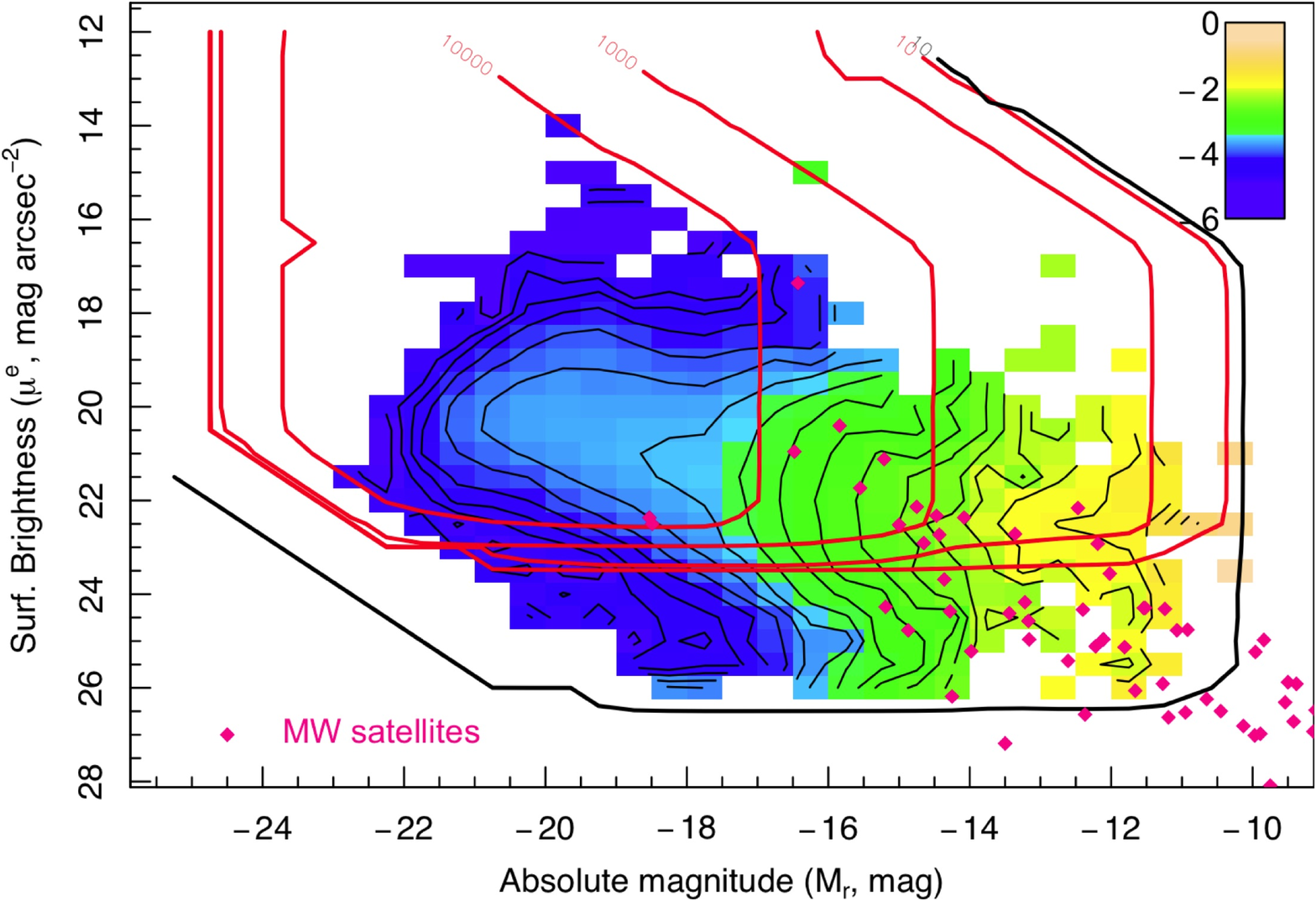}{SDriver-fig4x.eps}{The space-density of galaxies in the
  absolute magnitude surface brightness plane as sampled by GAMA
  (image and contours scaled logarithmically). The data points show
  the proof-of-existence presented by the local group satellites. The
  red lines show the iso-volume contours (in units of Mpc$^3$) and
  indicate the current GAMA selection boundaries. The black line shows
  the 10Mpc$^3$ selection boundary once VST KiDS observations are
  completed.}

\section{Stellar Masses and Galaxy Groups}

\vspace{-0.5cm}

\noindent
Two of the key GAMA catalogues produced to date are the stellar mass
catalogue (Taylor et al.~2013), and the galaxy group and pair
catalogue (Robotham et al.~2011). Fig.~3 shows the halo mass-redshift
plane sampled by the GAMA group catalogue compared to those derived
from SDSS and zCOSMOS. GAMA complements both surveys helping to bridge
the gap between the very near and very far Universe. Both the stellar
mass and group catalogues have been central in a number of defining
studies including the best current estimate of the local galaxy
stellar mass function (Baldry et al.~2012), the galaxy stellar-mass
function subdivided by morphological type (Kelvin et al.~2013), the
recalibration of the local mass-size relations (Lange et al.~2015),
the construction of filament and tendril catalogues (Alpaslan et
al.~2014a,b), the identification of local group analogues (Robotham et
al.~2014), the measurement of the low-z merger rate (Robotham et
al.~2014), and the propensity for major and minor mergers to
induce/suppress star-formation (Davies et al.~2015). In the near
future the group catalogue will be used to provide the definitive
measurement of the halo mass function; a key test of the dark-matter
paradigm with the potential to constrain the particle mass to $>$ keV
scales. High mass compact groups at intermediate redshifts provide the
optimal gravitational probes of the very high-z Universe, and
guaranteed-time observations of selected GAMA groups are planned with
the James Webb Space Telescope.

One current limitation of GAMA is its dependency on SDSS for the input
catalogue definition, introducing a bright surface brightness
selection limit $\mu_{\rm lim} \sim 23$ mag/sq arcsec. Fig.~4 shows
the space-density of galaxies in the luminosity-surface brightness
plane (unpublished). This plane is critical for managing and assessing
the selection boundaries dictated by the flux limit, surface
brightness limit, sky smoothing, and spatial sampling (see Driver 1999
and Driver et al.~2005 for details). The red lines on Fig.~4 indicate
iso-volume contours outside of which we cannot be certain that our
sample is complete. We see from Fig.~4 that GAMA (and similarly for
all other local surveys) hits the selection boundaries at the
giant-dwarf boundary (i.e., ballpark LMC luminosity/mass). In fact
only a very small portion of the parameter space over which galaxies
are known to exist is well sampled in a cosmologically representative
manner. Extending these boundaries is critical. VST KiDS is currently
surveying the GAMA regions and will extend the low surface brightness
boundary to $\mu_{\rm lim} \sim 26$ mag/sq arcsec (Fig.~4, black
line), very much opening our observing window into the luminous low
surface brightness Universe. Extending to very low luminosities/masses
is however trickier, and one of the goals of the next generation WAVES
survey (Driver et al.~2015).

We finish by once again highlighting the remarkable panchromatic
coverage of GAMA. Fig.~5 shows an example of the 21-band photometry
for just one of the 250k galaxies in our sample, along with a MAGPHYS
fit to the data. In the near future the GAMA fields will be surveyed
with the Australian Square Kilometer Array providing crucial insight
into the neutral gas domain and continuum emission from $\sim$ 20-40cm.

\section{Summary}

\vspace{-0.5cm}

\noindent
GAMA is now reaching maturity by providing high quality data products
for direct empirical studies, from which well defined sample can be
drawn and against which numerical models can be tested.  Moreover GAMA
is now poised to move forward into new territory pushing the selection
boundaries back in a comprehensive, complete and systematic
manner. Data releases (DR1, DR2 and PDR) are available via
http://www.gama-survey.org or feel free to contact us via
gama@eso.org.

\articlefigure{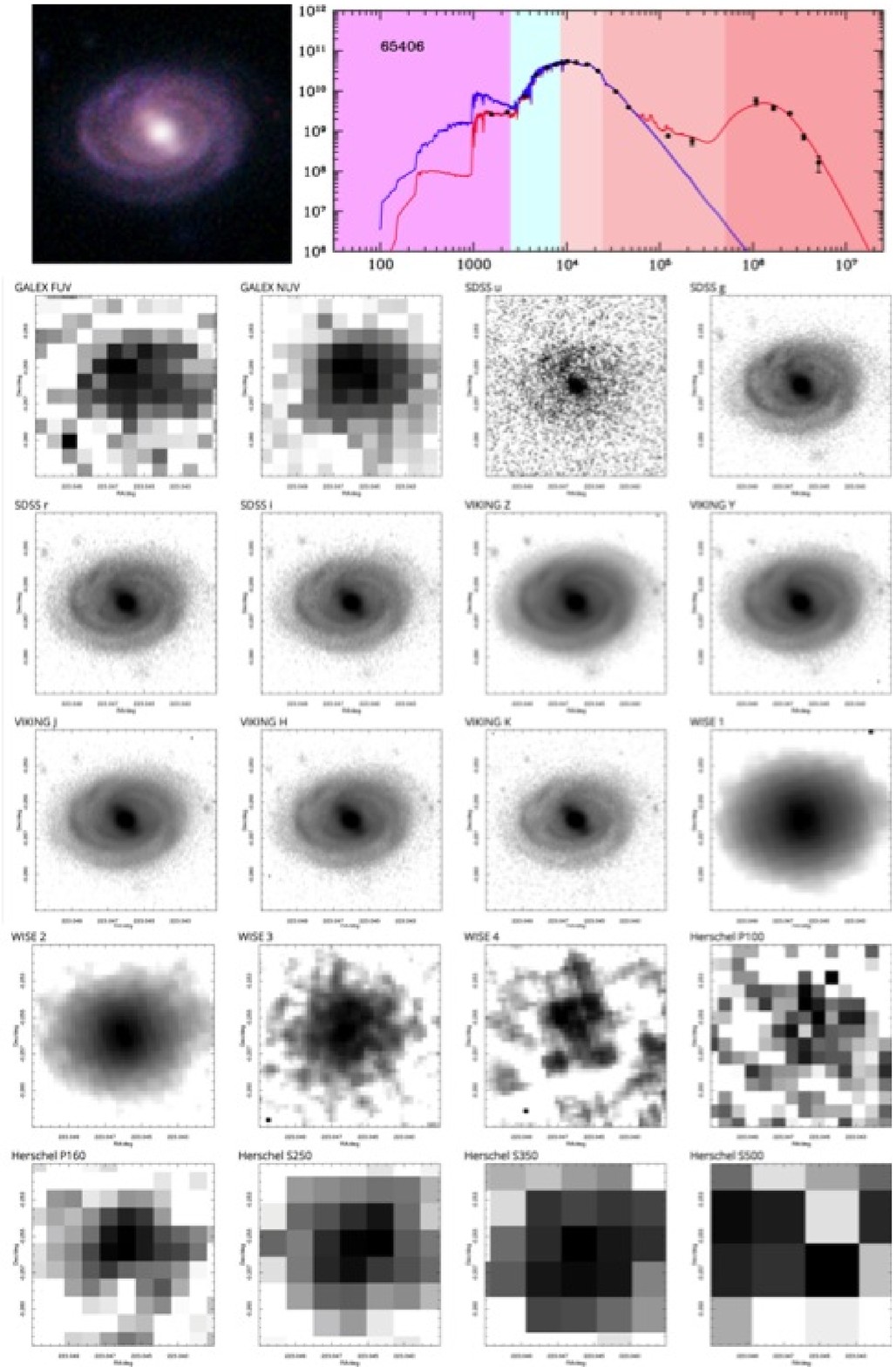}{SDriver-fig5x.eps}{Available panchromatic data for G65406 (SDSS z not shown) including MAGPHYS analysis (top right) and optical/near-IR $giH$ colour image (top left).}

\references

\vspace{-0.5cm}

\noindent
Alpaslan M., et al., 2014a, MNRAS, 440, 106
$\bullet$ Alpaslan M., et al., 2014b, MNRAS, 438,177
$\bullet$ Baldry I.K., et al., 2010, MNRAS, 404, 86
$\bullet$ Baldry I.K., et al., 2012, MNRAS, 421, 621
$\bullet$ Davies L., et al., 2015, MNRAS, submitted
$\bullet$ Driver S.P., et al., 2011, MNRAS, 413, 971
$\bullet$ Driver S.P., 1999, ApJ, 526, 69
$\bullet$ Driver S.P., 2005, MNRAS, 360, 81
$\bullet$ Driver S.P., 2015, ASSP, in press
$\bullet$ Lange R., et al., 2015, MNRAS, 447, 2603
$\bullet$ Liske J., et al., 2015, MNRAS, in press
$\bullet$ Hopkins A., et al., 2013, 430, 2047
$\bullet$ Kelvin L., et al., 2014, MNRAS, 444, 1647
$\bullet$ Robotham A.S.G., et al., 2010, PASA, 27, 76
$\bullet$ Robotham A.S.G., et al., 2011, MNRAS, 416, 2640
$\bullet$ Robotham A.S.G., et al., 2012, MNRAS, 416, 2640
$\bullet$ Robotham A.S.G., et al., 2013, MNRAS, 416, 2640
$\bullet$ Taylor E. et al., et al., 2013, MNRAS, 418, 1587

\noindent
\acknowledgements 
GAMA is a joint European-Australasian project based
around a spectroscopic campaign using the Anglo-Australian
telescope. The GAMA input catalogue is based on data taken from the
Sloan Digital Sky Survey and the UKIRT Infrared Deep Sky
Survey. 
\end{document}